\documentclass[12pt,preprint]{aastex}

\long\def\***#1{{\scshape ***#1***}}



\newcommand{\etal}{{\em et al.}}            

\shorttitle{Cen A discontinuity}
\shortauthors{Kraft \etal}

\begin{document}

\title{Evidence for Non-Hydrostatic Gas Motions in the Hot ISM of Centaurus A}
\author{R. P. Kraft\altaffilmark{1}, M. J. Hardcastle\altaffilmark{2}, G. R. Sivakoff\altaffilmark{3}, A. Jord{\'a}n\altaffilmark{1}, P. E. J. Nulsen\altaffilmark{1}, M. Birkinshaw\altaffilmark{4}, W. R. Forman\altaffilmark{1}, C. Jones\altaffilmark{1}, D. M. Worrall\altaffilmark{4}, J. H. Croston\altaffilmark{2}, D. A. Evans\altaffilmark{1}, S. Raychaudhury\altaffilmark{5}, S. S. Murray\altaffilmark{1}, N. J. Brassington\altaffilmark{1}, J. L. Goodger\altaffilmark{2}, W. E. Harris\altaffilmark{6}, A. M. Juett\altaffilmark{7}, C. L. Sarazin\altaffilmark{7}, K. A. Woodley\altaffilmark{6}}

\altaffiltext{1}{Harvard/Smithsonian Center for Astrophysics, 60 Garden St., MS-67, Cambridge, MA 02138}
\altaffiltext{2}{University of Hertfordshire, School of Physics, Astronomy, and Mathematics, Hatfield AL10 9AB, UK}
\altaffiltext{3}{The Ohio State University, Department of Astronomy, 4055 McPherson Laboratory, 140 W. 18th Ave., Columbus, OH 43210}
\altaffiltext{4}{University of Bristol, Department of Physics, Tyndall Avenue, Bristol BS8 ITL, UK}
\altaffiltext{5}{University of Birmingham, School of Physics and Astronomy, Edgebaston, Birmingham, B15 2TT, UK}
\altaffiltext{6}{McMaster University, Department of Physics and Astronomy, Hamilton, ON L8S 4M1, Canada}
\altaffiltext{7}{University of Virginia, Department of Astronomy, P. O. Box 400325, Charlottesville, VA 22904}

\begin{abstract}

We present preliminary results from a deep (600 ks) {\em Chandra} observation 
of the hot interstellar medium of the nearby early-type galaxy Centaurus A
(Cen A).  We find a surface brightness discontinuity in the
gas $\sim$3.5 kpc from the nucleus spanning a 120$^\circ$ arc.
The temperature of the gas is 0.60$\pm$0.05 and 0.68$\pm$0.10 keV, interior and exterior to
the discontinuity, respectively.
The elemental abundance is poorly constrained by the
spectral fits, but if the abundance is constant across the discontinuity,
there is a factor of 2.3$\pm$0.4 pressure jump across the discontinuity.
This would imply that the gas is moving at 470$\pm$100 km s$^{-1}$, or
Mach 1.0$\pm$0.2 (1.2$\pm$0.2) relative to the sound speed of the gas external (internal)
to the discontinuity.
Alternatively, pressure balance could be maintained if there is a large (factor of $\sim$7) 
discontinuity in the elemental abundance.
We suggest that the observed discontinuity is the result of non-hydrostatic
motion of the gas core (i.e. sloshing) due to the recent merger.
In this situation, both gas motions and abundance gradients are
important in the visibility of the discontinuity.
Cen A is in the late stages of merging with a small late-type galaxy,
and a large discontinuity in density and abundance across a short distance demonstrates
that the gas of the two galaxies remains poorly mixed even several hundred 
million years after the merger.  
The pressure discontinuity may have had a profound influence on
the temporal evolution of the kpc-scale jet.  The jet could have decollimated
crossing the discontinuity and thereby forming the northeast radio lobe.
\end{abstract}


\section{Introduction}

{\em Chandra} has revealed the complex relationship between the lobes
of radio galaxies and the hot gas of galaxies, groups, and clusters in
a large number of systems.
The heating of the gas induced by the inflation of the lobes, either
subsonic or supersonic, plays an important role in the energy budget
of the ISM of galaxies and the ICM of clusters.   Lobe inflation also
plays a key role in mixing the low entropy gas in the cores with the
high entropy gas in the halo \citep{roe07}.
In addition, the relatively low inertia of the radio lobes means that they
are easily influenced by subsonic or transonic motions of the 
ambient medium once the early, highly supersonic expansion phase has ended \citep{rey01}.
The temporal evolution of radio lobes could be dramatically altered in
systems that have undergone recent mergers.  
Observationally, there are at least two examples in which ongoing galaxy
mergers have dramatically influenced the evolution of radio
bubbles \citep{mjh07b,wor07}.
In {\it Chandra} observations both of the Antennae, the nearest
early-stage merger \citep{baldi06}, and of more distant merging
galaxies \citep{brass06b} we see large spatial variations
of gas temperature, density and elemental abundance as the hot ISM
of the merging galaxies begins to coalesce.
The dissipative processes that smooth out these gradients are
poorly understood, but transonic gas motions and temperature/pressure variations could play an
important role in the time evolution of relatively weak nuclear
outbursts.

Centaurus A (Cen A, NGC 5128) is the nearest radio galaxy and the nearest large, 
unobscured, early-type galaxy (D$_L$=3.7 Mpc, $M_B=$-21.2, see \citet{duf79}).
It is the dominant member of a poor group,
and the prototypical example of several astrophysically interesting
categories, including low-luminosity (FR I) radio galaxies,
early-type galaxies, and late-stage mergers.
These make it the ideal target for the study of the dynamics of the merger
process and its effects on radio bubbles.
Six 100 ks observations were made with {\em Chandra}/ACIS-I in AO-8
to study the X-ray jet \citep{kra02,mjh03,mjh07b,wor08},
the interaction of the radio lobes with the ambient gas \citep{kra03}, the X-ray binary
population \citep{kra01,voss06,jor07,siv08}, and the hot gas \citep{kar02,kra03}.

In this Letter, we present preliminary results from our deep 
AO-8 {\em Chandra} observation of the ISM of Cen A and its interaction with the northeast
radio lobe.  In particular, we report the discovery of a surface brightness discontinuity 
in the gas $\sim$3.2 kpc from the nucleus that spans a 120$^\circ$ arc.
We conclude that this discontinuity and associated gas motions
may be responsible for the observed radio morphology of the northeast radio lobe.
We adopt a distance of 3.7 Mpc to Cen A, which is the mean of distance measures 
from five independent indicators (TRGB, Mira variables, SBF, PNLF, Cepheids, 
summarized in Section 6 of \citet{fer07}).
At this distance, 1$''$=17.9 pc and 1$'$=1.076 kpc.
All uncertainties are at 90\% confidence for one parameter of interest
unless otherwise stated, and all coordinates are J2000.
All spectral fits include absorption ($N_H$=8.41$\times$10$^{20}$ cm$^{-2}$)
 by gas in our Galaxy \citep{dic90}.

\section{Data Analysis}

Cen A has been observed ten times with {\em Chandra}/ACIS.
All the data were reprocessed to apply the most up to date gain and CTI correction, 
and the event files were filtered to remove events at node boundaries.  Standard
ASCA grade filtering (i.e. event grades 0,2,3,4,6) was applied
to the data.  A lightcurve was created in the 5-10 keV band after 
removal of point sources to search for periods of background flaring.  
Intervals where the background rate was more than 3$\sigma$ above the
mean were removed, leaving $\sim$623 ks and $\sim$96 ks of good time for
the ACIS-I and ACIS-S observations, respectively.
The observations were co-aligned by matching the X-ray positions of
$\sim$200 X-ray binaries and other point sources (about half having $>$100
counts in the 0.5--7.0 keV band within a single observation) \citep{jor07}. 
Full details of the data processing and alignment will
be presented in a future publication.  Given the narrow scope of this Letter,
we use only the ACIS-I data (roughly 85\% of the observation time)
to avoid systematic uncertainties due to different backgrounds
and spectral responses between the ACIS-S and ACIS-I detectors.

\section{Results}

\begin{figure}
\plotone{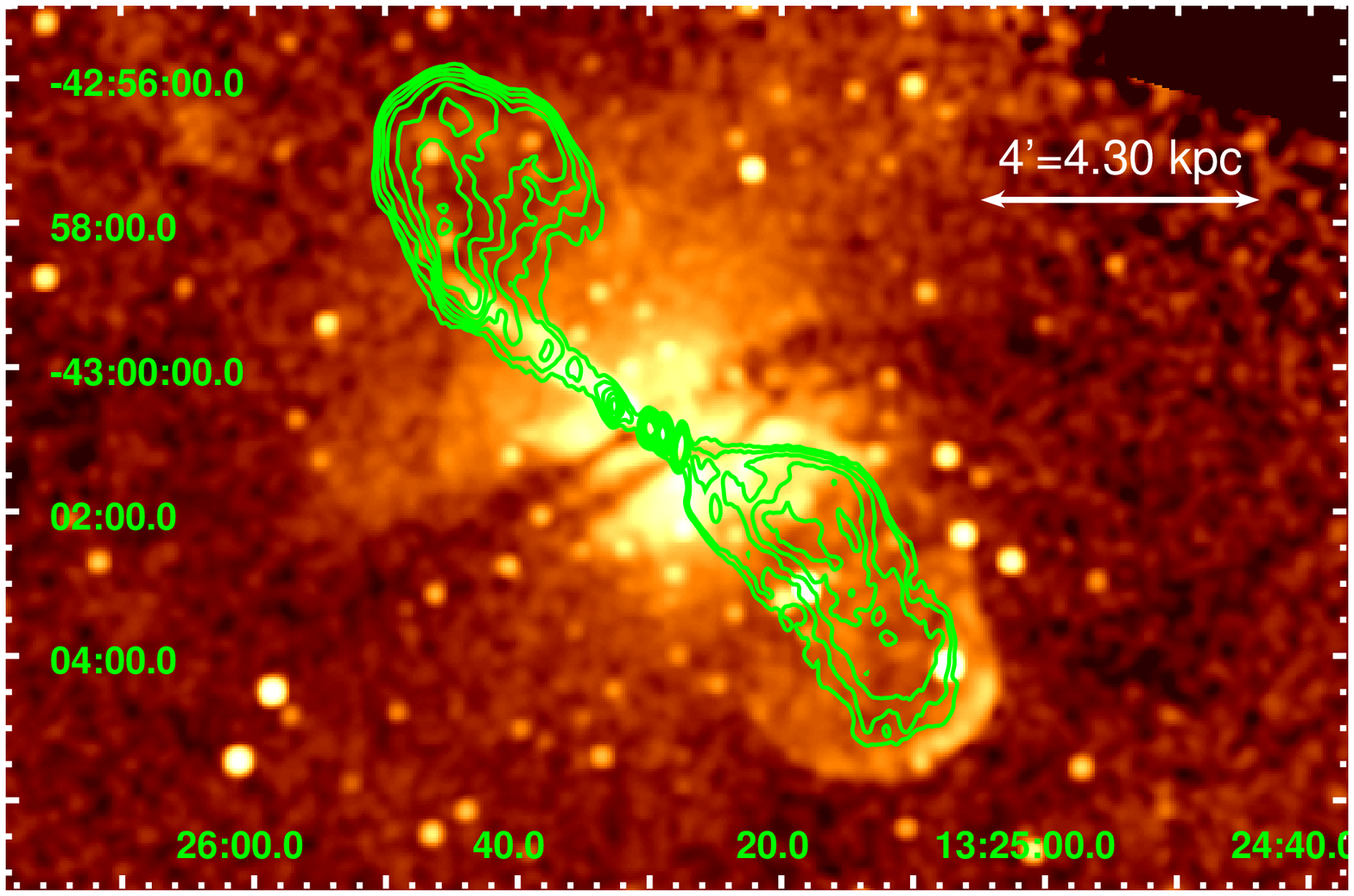}
\hspace{0.5in}
\caption{Gaussian smoothed, exposure corrected {\em Chandra}/ACIS-I image 
of Cen A in the 0.5-1.0 keV band with 5 GHz radio contours (6$''$ resolution) overlaid.
The color stretch has been selected to enhance the appearance of the X-ray surface
brightness discontinuity to the northeast of the nucleus.}\label{cenatotal}
\end{figure}

\begin{figure}
\plotone{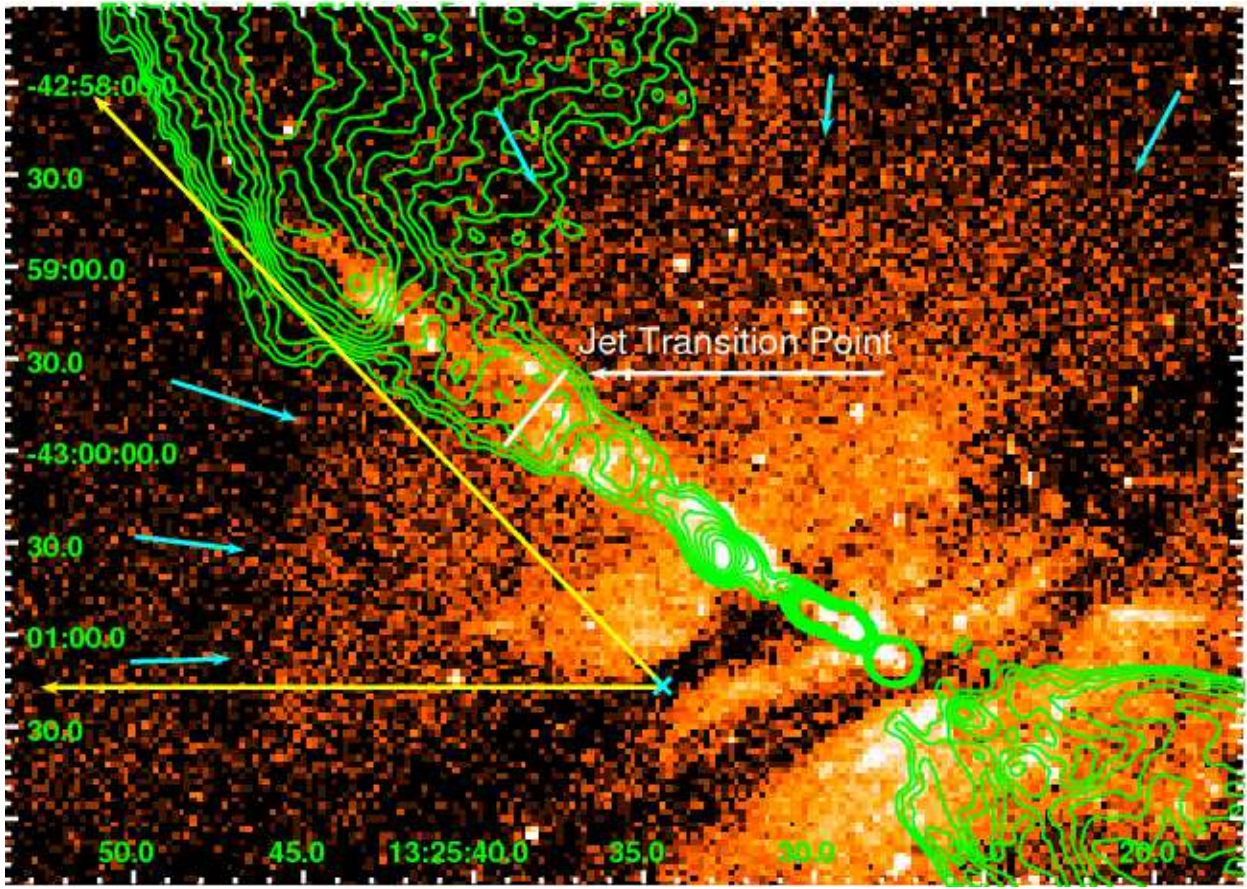}
\caption{Raw {\em Chandra}/ACIS-I image 
of Cen A in the 0.5-1.0 keV band.  The light blue arrows denote the approximate
position of the surface brightness discontinuity in the gas.
The light blue cross and yellow lines denote the wedge in which the surface
brightness profile in Figure~\ref{sbprof} was created.  The approximate
position of the transition region of the jet (where the width of the X-ray
jet narrows) is also labeled.}\label{discont}
\end{figure}

A Gaussian smoothed, exposure corrected {\em Chandra}/ACIS-I image of Cen A
in the 0.5-1.0 keV band with 5 GHz radio contours (6$''$ resolution) overlaid is shown in
Figure~\ref{cenatotal}.  
One of the most striking features of Figure~\ref{cenatotal} is the
surface brightness discontinuity which lies $\sim$3.2$'$ (3.54 kpc)
to the east and north of the nucleus.  
Figure~\ref{discont} contains an unsmoothed X-ray image in the same
band in the vicinity of the discontinuity (denoted by the blue arrows).

\begin{figure}
\plotone{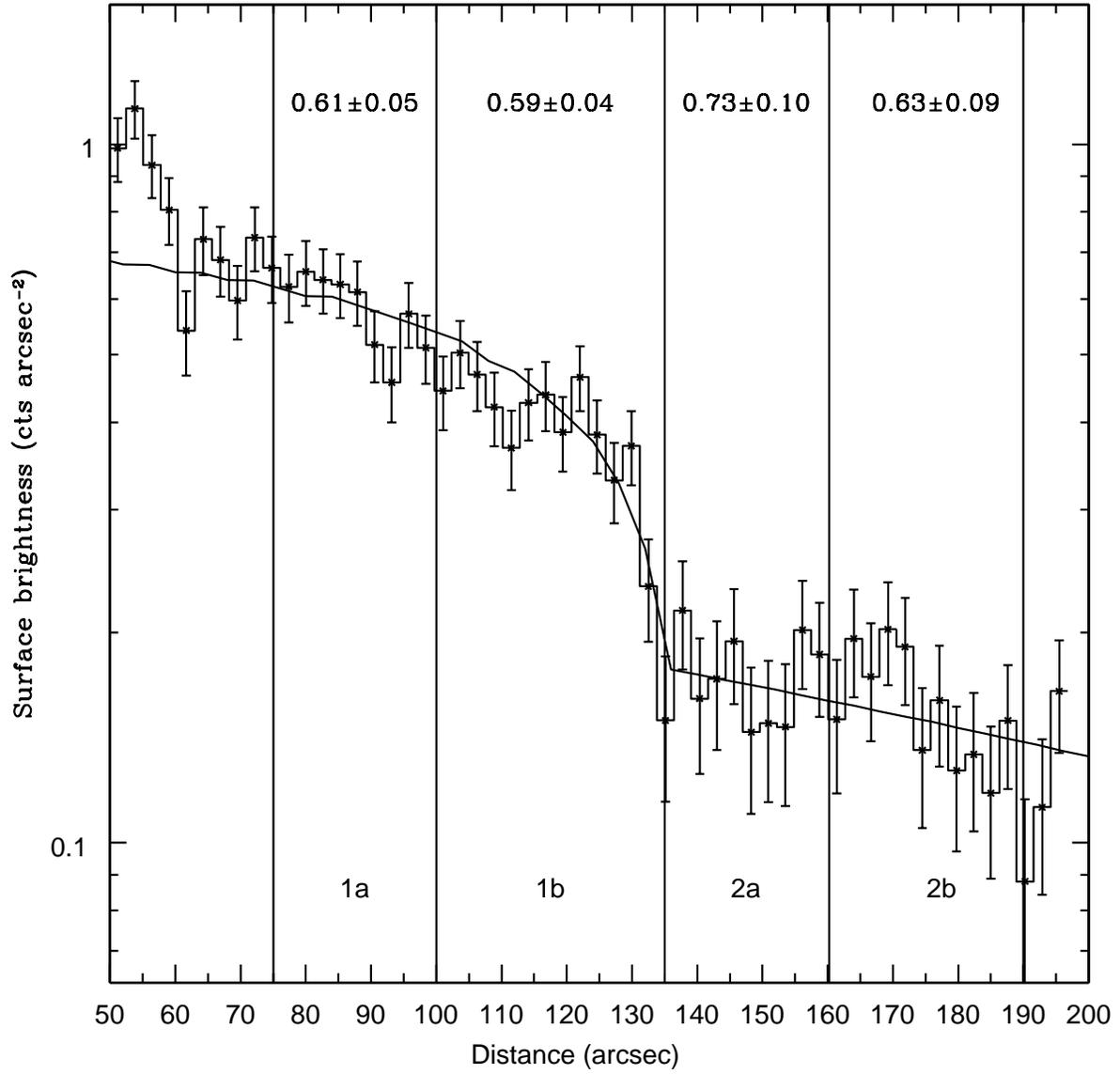}
\caption{Surface brightness profile of the hot ISM in a 50$^\circ$
wedge (shown in Figure~\ref{discont}) in the 0.5-1.0 keV band.
The continuous curve is the best fit surface brightness profile model consisting
of a uniform density sphere out to radius 134$''$ and a beta model
profile at larger radii.}\label{sbprof}
\end{figure}

We extracted a surface brightness profile, shown in Figure~\ref{sbprof},
across the discontinuity in a 50$^\circ$ pie slice region centered at $\alpha$=13:25:34.445,
$\delta$=-43:01:16.84.  
The vertex of this region was located $\sim$1.3$'$ east of the nucleus,
so that the radius of curvature of the discontinuity
matches that of the annular regions, simplifying the deprojection of
surface brightness to density.
We fitted spectra to four radial regions (1a, 1b, 2a and 2b) 
of the galaxy gas.  Background was determined from a distant region of the combined image
free from emission from Cen A.  Absorbed APEC models were fitted to each spectrum,
with the absorption fixed at the Galactic value, and the abundance
allowed to vary freely.  The two spectra on the interior of the discontinuity
(1a and 1b) contained an additional component, the contribution to the emission
from the overlying gas.  The best fit temperatures and 90\% uncertainties
are shown in Table~\ref{tempfit}.  In all cases, the elemental abundance was
unconstrained.  At these temperatures, the emission is line dominated
so that the normalization can be traded off with the abundance.  
The only formal statistical constraint from each fit was that the elemental abundance,
$Z$, is $>$0.1 at 90\% confidence in the two regions interior to the discontinuity
(regions 1a and 1b).

\section{Interpretation}

We deprojected the surface brightness profile assuming a uniform
gas density out to $R_{discont}$ (134$''$), and a beta-model gas
density profile with core radius
(fixed) $r_0$=200$''$ beyond $R_{discont}$ centered at a point $\sim$1.3$'$ east of
the nucleus (see Figure~\ref{discont}).  
Our general conclusions are not sensitive to
these model assumptions.  The temperatures of the interior and exterior gas
were taken to be 0.6 and 0.68 keV, respectively, and represent the average
values of regions 1a+1b and 2a+2b.
The elemental abundance was unconstrained in the spectral fits, so we
initially assumed uniform abundance ($Z$=1.0) on both sides of the
discontinuity.  The best fit surface brightness profile using the density profile
described above is shown overplotted
on Figure~\ref{sbprof}.  We find the hydrogen density interior
and exterior to the discontinuity to be 3.60$\times$10$^{-3}$ and 1.38$\times$10$^{-3}$
cm$^{-3}$, respectively.
The density ratio is more than double the ratio of the temperature differences and implies
that the interior gas is greatly overpressurized relative to the exterior
gas.  In this case, the interior gas should be expanding at roughly
its internal sound speed and the pressure gradient would dissipate in roughly a sound
crossing time.

One possible interpretation of this phenomenon is that the surface brightness
discontinuity and density jump are the
result of `sloshing', that is non-hydrostatic motions of the central
dense regions of the ISM due to the merger \citep{mar07}.  In this scenario,
the gas is moving to the northeast at transonic velocity and is oscillating,
perhaps non-radially, in the gravitational potential of the galaxy.
This model has been invoked to explain the presence of similar features
observed in clusters of galaxies.
Hydrodynamic simulations of this phenomenon
for galaxy clusters indicate that the non-hydrostatic oscillation of the 
core can be quite long lived, typically several Gyrs.  The length scales
are much smaller in Cen A than in galaxy clusters, but the sound speed of the
gas is also much lower.  The presence of the `sloshing' gas is
entirely consistent with the time since the merger (several hundred
Myr - see \citet{isr98}).
The ratio of the gas pressures interior and exterior to the discontinuity
is 2.3$\pm$0.4.  If this ratio represents the ratio of pressures at the
stagnation point and free stream region, the gas interior to the discontinuity
is moving to the northeast at approximately 470 km s$^{-1}$, or with Mach number
1.0$\pm$0.2 (1.2$\pm$0.2) relative to the gas exterior (interior) to the 
discontinuity \citep{lan50}.

In this analysis we assumed, however, that the elemental abundance is constant 
across the discontinuity.  If there is a large, discontinuous jump in the 
elemental abundance, the gas on the two sides of the surface brightness discontinuity 
could be at or near pressure equilibrium.
This would imply that the elemental abundance of the material interior to
the discontinuity was several times larger than that exterior to the
discontinuity, and that the ratio of the gas densities is the reciprocal
of the temperature ratio.
In this model, the surface brightness discontinuity is a relatively stable,
long lasting contact discontinuity between two fluids.
The cooling function, $\Lambda$, of an optically thin
plasma scales nearly linearly with elemental abundance at the
temperatures ($<$0.75 keV) and abundances relevant to the hot ISM of Cen A.
The difference from a purely linear relationship is less than 5\%
for $Z>$0.1, and since $n_H\sim\Lambda^{-0.5}$, these
differences are negligible.
The elemental abundances of the gas interior and exterior to the
discontinuity would therefore have to differ by a factor of $\sim$7
for the two fluids to be in pressure equilibrium.
Thus, if the exterior gas is relatively metal poor ($Z_{ext}$=0.1-0.3), the
gas interior to the discontinuity would have to have Solar or super-Solar
abundance ($Z_{int}$=0.7-2.1) for the pressure to be continuous.

Such a low abundance of the gas exterior to the edge is plausible.  
Recent HI observations of a sample of relatively isolated elliptical and
lenticular galaxies demonstrates that
there is considerable cool gas residing in their halos \citep{oost07}.
If such primordial cold gas is continuously falling into Cen A, the
elemental abundance of the outer parts of the hot ISM could be quite low.
Arecibo measurements of HI around early-type galaxies in and around
the Virgo cluster have found little cold gas \citep{alig07}.  In dense
environments, HI is likely to be stripped, but this is not expected in a poor
environment like the Cen A group \citep{isr98}.
A recent X-ray study of the hot ISM of the gas poor, isolated early-type galaxy NGC 4697
found $Z\sim$0.25$Z_\odot$ \citep{irwin07}, consistent with this picture.
There have been several measurements of the elemental abundance of the stellar halo population
of Cen A.  \citet{har00} reported that the average abundance of the stars in the halo
of Cen A ($\sim$20-30 kpc from the nucleus) is $Z\sim$0.37$Z_\odot$.  After subtraction of the large-scale
halo contribution, \citet{har02} found that the the metallicity of stars closer to the
nucleus ($\sim$8 kpc) was somewhat more metal rich, with $Z\sim$0.65$Z_\odot$.
Suzaku measurements of emission lines and elemental absorption edges
of the circumnuclear gas in Cen A suggest that the elemental abundance is
slightly super-Solar ($Z$=1.3) \citep{markz07}.  
However, Suzaku does not have the angular resolution to resolve any of the substructure
in the gas, the variable absorption, or the X-ray binary population.
Analysis of {\em Chandra} spectra of the gas in the inner 2 kpc of Cen A 
demonstrates that the gas temperature of the central region varies by more than a factor of
two, and there is a spatial variation in the elemental abundance \citep{brass06a}.
The temperature and elemental abundance of the central region based on
the VLP observations will be presented in a future publication (Kraft \etal~2008, in preparation).

In the absence of a precise measurement of the elemental abundances across the
discontinuity, we conclude that both phenomena (i.e. gas motions and
elemental abundance gradients) may be contributing to the
visibility of the surface brightness discontinuity.
The abundance profile in many early-type galaxies is sharply peaked
at the center, and if the central core is rapidly displaced off-center,
a sharp surface brightness discontinuity could result even at zero
velocity.  The inferred velocity of $\sim$470 km s$^{-1}$ should therefore 
be regarded as an upper limit.
The effect of projection is another source of systematic uncertainty in
this analysis.  Three dimensional simulations of sloshing in galaxy cluster mergers \citep{mar07}
show complex gas temperature and density variations
that would be difficult, if not impossible, to resolve via deprojection of surface
brightness profiles.

We consider the possibility that this surface brightness discontinuity is
a shock due to the inflation of radio lobe to be unlikely for several reasons.
First, it is not at all clear what is driving the shock.  The pressure is
larger on the interior, so the shock must be driven from the inside out.
The northeast radio lobe clearly lies outside the discontinuity, so the inflation
of the lobe cannot be responsible.  There is nothing that appears to be driving
the gas at supersonic velocity more or less radially from the nucleus.  Second,
the surface brightness and temperature profiles are wrong for a shock.
We should observe a sharp jump in the surface brightness and gas temperature just 
behind the shock.  This temperature jump, even for a high Mach number shock,
would easily be observable by {\em Chandra} in Cen A if present.  
Third, there is no corresponding feature on the opposite side of Cen A.
It is hard to imagine how the gas could be shock heated on only one side of the galaxy.

\section{Implications for Evolution of the Jet and Radio Lobes}

Did the discontinuity in the gas affect the temporal evolution of the jet
and northeast radio lobe?  The X-ray morphology of the jet shows a nearly
constant opening angle to a distance of 139$''$ from the nucleus,
where the X-ray jet narrows (see Figure~1 of \citet{mjh07b}).
It is at about this point that the jet opens into the northeast lobe in the radio band.
One possibility is that the point at which the X-ray jet narrows is
the position at which the jet crosses the X-ray surface brightness discontinuity.
This point lies interior to the discontinuity in projection since the jet
does not lie directly in the plane of the sky.
The sudden change in ambient gas density and pressure across the
discontinuity could explain the transition from jet to lobe.
\citet{nor88} modeled this jet to lobe transition in Cen A as the
result of the jet encountering a sudden {\it increase} in the external pressure
and the jet transitioning from supersonic to subsonic relative to
its internal sound speed.  They hypothesized that gas was infalling
on Cen A and driving a shock in toward the center the galaxy.
Our data suggest that it is, in fact, the opposite scenario that is occuring.
The motion of the gas interior to the discontinuity is creating
the pressure jump (decrease) that de-collimates the jet.  

The decollimation of the jet across the discontinuity may
account for some of the observed features of the X-ray jet.
There must clearly be a change in the efficiency of particle
acceleration at the point where the jet cross the
discontinuity.  The X-ray emission from the jet narrows and fades away
as the jet enters the lobe \citep{mjh07b}.  
The spectral index of the jet emission beyond the discontinuity
is considerably steeper than either the compact knots or
diffuse emission of the rest of the jet.  There is no obvious structure 
in the radio emission, however.  The radio jet both widens and brightens 
as it smoothly enters the northeast radio lobe.
Rapid expansion of the jet may lead to a reconfinement shock,
causing the jet to decelerate, and will certainly cause the
magnetic field strength in the jet to fall:  both of these could
reduce the efficiency of acceleration of ultra-relativistic
particles.

Can the change in physical conditions at the discontinuity be related to the
bending of the lobe toward the northwest?
Fundamentally, the bending of the X-ray jet into the northeast lobe is
likely the result of velocity and pressure gradients in the external gas,
although the low surface brightness of the gas and uncertainties due
to projection prevent us from making a definitive statement.
If the core is `sloshing' there will be
significant subsonic or transonic motions of the gas exterior to the
discontinuity.  This implies pressure variations of tens of percent to perhaps
as large as a factor of 2.
However, sharp X-ray surface brightness discontinuities have been observed along the north
and northwest boundaries of the northeast
lobe suggesting that it is expanding into the ISM supersonically \citep{kra07}.
We also know that the southwest radio lobe is expanding into the ISM highly
supersonically.  The similarity in size of the northeast and southwest lobes
demonstrates that they must be inflating at roughly equal rates, although
there are obvious bends in the jet in projection, so there may be some
difference in the sizes of the lobes.
A nearly transonic wind striking a radio lobe could cause it to inflate asymmetrically
if the lobe is overpressurized by at most a factor of a few relative to the ambient medium.

Finally we note that the position of the X-ray surface brightness discontinuity
perfectly overlaps the position of one of the optical shells seen in deep exposures
of Cen A \citep{mal83,gk84}.
These optical shells (in Cen A and other early-type galaxies)
are believed to be the result of phase wrapping
of the dynamically cold disk of the merging spiral galaxy in the gravitational potential of 
the more massive early-type galaxy \citep{quinn84}.
The motions of the stars in these shells are thought to be a small
perturbation in the larger scale potential.
If the association of the X-ray surface brightness discontinuity with the
stellar shell is not coincidental, the gas and some of the associated stars
must therefore be moving synchronously.
Interestingly, both HI and CO have been observed to be associated with
the optical shells of Cen A as well \citep{ds94,char00}.
The appearance of the optical shells may not be the result of phase-wrapping
but could be indicative of other dynamic stellar motions.

\section{Acknowledgements}

This work was supported by NASA grant GO7-8105X and the Royal Society.
We thank the anonymous referee for detailed comments that improved this paper.

\clearpage

\clearpage

\clearpage

\clearpage

\begin{table}
\begin{center}
\begin{tabular}{|c|c|}\hline
Region & Temperature (keV) \\ \hline\hline
 1a & 0.61$\pm$0.05 \\ \hline
 1b & 0.59$\pm$0.04 \\ \hline
 2a & 0.73$\pm$0.10 \\ \hline
 2b & 0.63$\pm$0.09 \\ \hline
\end{tabular}
\end{center}
\caption{Best fit temperatures and uncertainties (90\% confidence)
for two regions interior (1a and 1b) and two regions exterior (2a and 2b) to
the surface brightness discontinuity.}\label{tempfit}
\end{table}

\end{document}